\documentclass[conference]{IEEEtran}
\IEEEoverridecommandlockouts
\usepackage{cite}

\usepackage{tikz}
\newcommand\copyrighttext{%
  \footnotesize \textcopyright \the\year{} IEEE. Personal use of this material is permitted. Permission from IEEE must be obtained for all other uses, including reprinting/republishing this material for advertising or promotional purposes, collecting new collected works for resale or redistribution to servers or lists, or reuse of any copyrighted component of this work in other works.}

\newcommand\copyrightnotice{%
\begin{tikzpicture}[remember picture,overlay]
\node[anchor=south,yshift=10pt] at (current page.south) {\fbox{\parbox{\dimexpr0.75\textwidth-\fboxsep-\fboxrule\relax}{\copyrighttext}}};
\end{tikzpicture}%
}
\renewcommand\fbox{\fcolorbox{red}{white}}
\usepackage{url}
\usepackage{amsmath,amssymb,amsfonts}
\usepackage{algorithmic}
\usepackage{graphicx}
\usepackage{textcomp}
\usepackage{xcolor}
\usepackage{enumitem}
\def\BibTeX{{\rm B\kern-.05em{\sc i\kern-.025em b}\kern-.08em
    T\kern-.1667em\lower.7ex\hbox{E}\kern-.125emX}}
\begin{document}


\title{Enabling SSI-Compliant Use of EUDI Wallet Credentials through Trusted Execution Environment and Zero-Knowledge Proof
}

\author{\IEEEauthorblockN{Nacereddine Sitouah\IEEEauthorrefmark{1}, Francesco Bruschi \IEEEauthorrefmark{1},   Stefano De Cillis \IEEEauthorrefmark{2} }
\IEEEauthorblockA{\IEEEauthorrefmark{1}Department of Electronics, Information and Bioengineering - Polytechnic University of Milan, Milan, Italy \\  first.last@polimi.it}
\IEEEauthorblockA{\IEEEauthorrefmark{2}     stefanodecillis96@gmail.com} }


\maketitle
\copyrightnotice
\begin{abstract}
 The passing of the eIDAS amendment marks an important milestone for EU countries and changes how they must manage digital credentials for both public services and businesses. Italy has led in adopting eIDAS, first with CIE and SPID identity schemes, and now with the Italian Wallet (IO app) aligned to eIDAS 2.0.  Self-Sovereign Identity (SSI) is a decentralized model born from the success of Distributed Ledgers, giving individuals full control over their digital identity. The current eIDAS 2.0  and its implementation acts diverge from SSI principles, rendering the European Digital Identity Wallet (EUDIW) centralized and merely user‑centric,  prioritizing security and legal protection over true self‑sovereignty.
 This paper proposes an architecture that enables the use of IT Wallet credentials and services in an SSI-compliant environment through Trusted Execution Environments and Zero-Knowledge Proofs.
\end{abstract}


\section{Introduction}\label{Introduction}
Digital Identity Management Systems (IDMS) provide individuals with complex platforms and services to seamlessly have an online presence, encompassing operations that allow the identification, authentication and authorization of individuals' access privileges to digital resources \cite{laurent2015digital}. A robust and sound IDMS is key to trust, security, and a seamless digital experience for clients and businesses

Decentralized identity represents a modern approach to identity management. Before the rise of Distributed Ledger Technologies and Turing-complete blockchain networks---i.e  Ethereum and the Ethereum Virtual Machine (EVM)---digital identities were primarily managed by centralized authorities and identity providers. Over time, systems evolved to improve scalability and convenience through federated management and user-centric approaches, granting users certain capabilities, mainly consent. Nevertheless, reliance on trusted providers remained inherent assumptions or requirements \cite{zwitter2020digital}.

\begin{figure}[!t]
\centerline{\includegraphics[width=9.3cm]{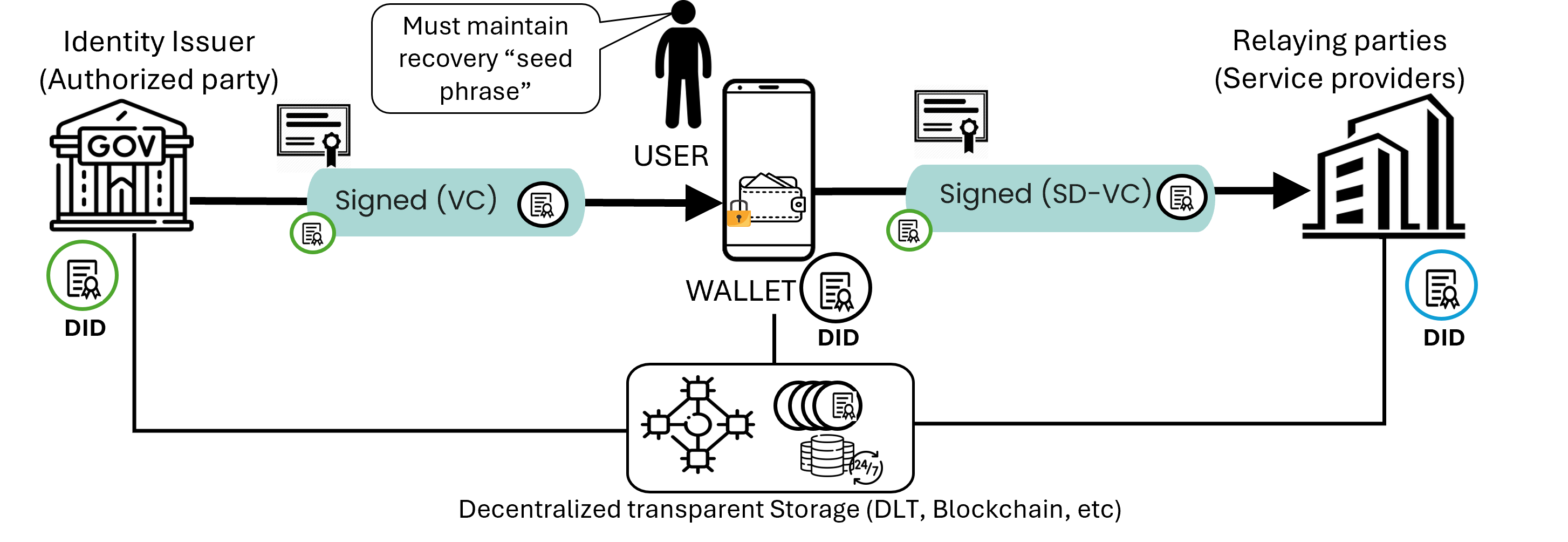}}
\caption{ Overview of a typical SSI interaction}
\label{Overview of a typically SSI interaction}
\end{figure}

Self-Sovereign Identity (SSI)---or decentralized identity---defer from previous identity models, using the decentralized nature of blockchain networks to put the users in control of their identity information \cite{zwitter2020digital,1, 9858139}. It is typically built around Decentralized Identifiers (DIDs) which are unique, user-controlled identifiers that can be resolved to metadata like public keys, and Verifiable Credentials (VCs), digitally signed attestations issued by trusted entities or credential issuers (e.g., universities, governments) about a subject. These credentials are highly available, they can be selectively disclosed and always cryptographically verified by relaying parties, enabling privacy-preserving authentication and sovereignty in digital interactions---as shown in Fig. \ref{Overview of a typically SSI interaction}.
 The EU opted for a regulation that creates a single framework for electronic IDentification (eID) and trust services that promotes interoperability across Member States, ensures safe electronic interactions between businesses, while improving efficiency and service quality.  eIDAS\footnote{Regulation (EU) No 910/2014  of the European Parliament and of the Council of 23 July 2014 on electronic identification and trust services for electronic transactions in the internal market and repealing Directive 1999/93/EC ({http://data.europa.eu/eli/reg/2014/910/oj}).} stands for electronic IDentification, Authentication and Trust Services, communicated first by the European Commission (EC) in June 2012\footnote{COM (2012) 238 : Proposal for a REGULATION OF THE EUROPEAN PARLIAMENT AND OF THE COUNCIL on electronic identification and trust services for electronic transactions in the internal market ({https://eur-lex.europa.eu/procedure/EN/2012\_146}).} with the goal of securing seamless electronic interactions between citizens, businesses, and authorities to improve the effectiveness of both public and private online services and e-commerce in the EU. eIDAS 1.0 struggled with regulatory rigidity and inconsistencies among Member States, which hindered interoperability and EU-wide integration. Its complexity and lack of flexibility also limited private sector inclusion and broader eID adoption.

eIDAS 2.0  aims to resolve these issues by introducing the European Digital Identity Wallet (EUDIW) promoting cross-border trust, greater flexibility, and stronger inclusion of private services within a unified EU framework\cite{schwalm2022eidas}. The framework envisions both public and private types of wallets, enabling Member States and trusted private providers to offer interoperable solutions that adhere to common EU standards.

Three key challenges prevent private EUDIW from achieving SSI compliance:
\begin{enumerate}
    \item Authority centralization: status endpoints rely on Public Key Infrastructures (PKIs), requiring providers and trust anchors to run services that return the status of trust‑marks, credentials, or wallet instances
    
    \item Exclusivity of Identification Data: in the EUDIW, the regulation explicitly defines identification data as government-issued Personal Identification Data (PID) that serve as the legal basis for authentication. 
   This means that only PIDs can be legally used for authentication; no other identifiers are permitted. 
    Consequently, the anonymization supported by eIDAS does not extend to wallet providers, as they can identify users at all times. 
    \item Relaying parties registration: verifiers of PIDs, Qualified Electronic Attestations of Attributes (QEAAs) and any other attestation must be registered, which limits the usability and potential use cases of the Wallet.
\end{enumerate}

These challenges make a private EUDIW incompatible with SSI, as authentication is limited to PIDs and pseudonyms remain traceable under centralization.
This study addresses three core research questions: how EUDIW credentials can be integrated into an SSI‑compliant ecosystem, how identity holders can autonomously generate verifiable proofs of their credentials, and how these proofs can be made fully SSI‑compatible within the system.

\section{preliminaries}
\subsection{IO IT Wallet and eIDAS 2.0}
The Italian “IO” app\footnote{{ IT-Wallet System --- app IO ({https://ioapp.it/documenti-su-io})}}---often referred to as the Italian Wallet (IT-Wallet)---is the official digital wallet and public services app developed by the Italian government, as part of Italy’s national digital identity ecosystem  and the country’s implementation of the EUDIW initiative under eIDAS 2.0. In its current version, the Wallet's functions are integrated into "IO" app in order to  allows citizens to request and store digital versions of Driver's license, Health card and the European disability card. It grants these digital documents the same legal value of physical cards. IT-Wallet authenticates its users using the existing eID schemes: SPID (Sistema Pubblico di Identità Digitale) and CIE (Carta d’Identità Elettronica).

\begin{figure}[htbp]
\centerline{\includegraphics[width=9cm]{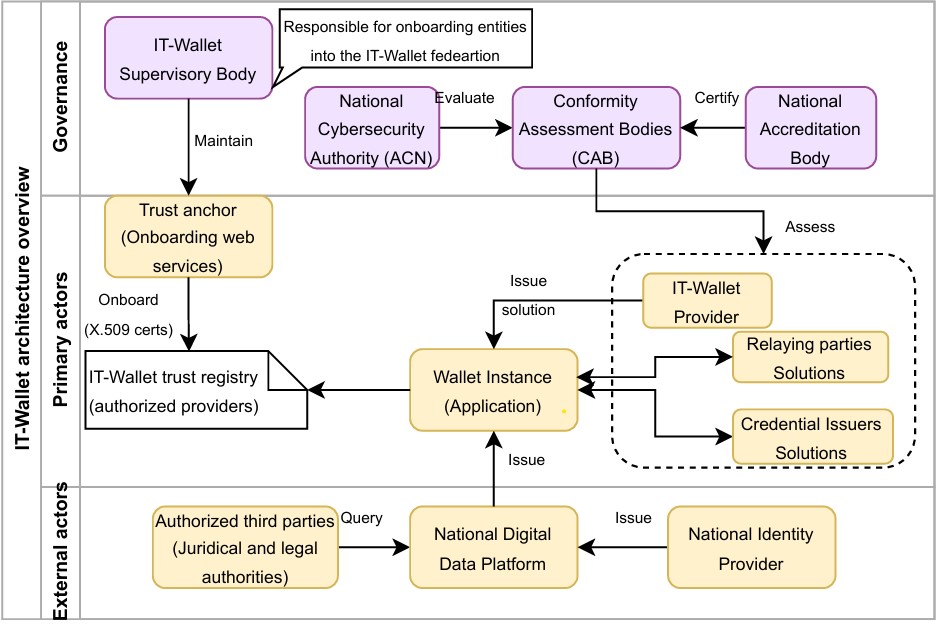}}
\caption{ IT-Wallet trust federation architecture: actors interaction \cite{agid2024itwallet}}
\label{IT-Wallet trust federation architecture: actors interaction}
\end{figure}

IT-Wallet relies on a Trust Federation (TF) architecture, a multi-layered architecture where governance bodies establish and maintain the trust infrastructure across Italy and the EU. This federation is based on OpenID Connect () Federation \cite{openid2024federation1_0} and adopted to meet eIDAS 2.0 Architecture Reference Framework guidelines---as illustrated in Fig. \ref{IT-Wallet trust federation architecture: actors interaction}.  

The verification of a PID or any credential requires validating the entire trust chain back to the trust anchor established by the Member State.  
Each participant in the federation possesses an Entity Configuration, a signed metadata document that describes it and is issued by a certification authority to attest to its compliance.
This configuration includes one or more trust-mark status endpoints, which typically  return a signed JSON response indicating the current validity status of the entity configuration; a silent or negative response signifies that the entity’s certification is invalid. A credential is considered valid only if all trust chain signatures can be successfully verified back to the system’s trust anchor and each trust-mark status endpoint confirms a positive (“active”) status.
\subsection{Verifiable Computation technologies}
\subsubsection{Zero knowledge proof (ZKP)}
 a ZK protocol allows one party---the prover---to convince another---the verifier---that a statement about some secret data is true without revealing the data itself. In the context of proof of computation, this means the verifier can check that a computation was carried out meeting certain conditions while remaining completely oblivious to the underlying or intermediate steps. ZK proofs thus preserve the confidentiality of inputs while still providing assurance of correctness. They establish a trust model in which the verifier relies not on the prover’s honesty, but solely on the mathematical soundness of the proof. However, constructing ZK proofs often requires defining a large set of deterministic constraints, which can impose substantial computational and implementation overhead \cite{zhang2025zkvc}.
 
\subsubsection{Trusted Execution Environment (TEE)}
A TEE is generally a reserved secure area within a hardware processing unit that can execute code in isolation from the rest of the system. This isolation safeguards against compromised OS and vulnerable software, ensuring that data within a TEE remain encrypted and inaccessible to unauthorized actors. 
Approaches like AMD SEV and Intel SGX/TDX \cite{b1} enable confidential workloads, supporting  sensitive tasks such as secure authentication, payments, and biometrics. Secure cloud computing leverages TEEs to keep client data confidential, shielding it from malicious actors, other tenants, and even cloud providers.

A TEE uses attestations to establish trust in the software it runs by verifying authenticity, integrity, and secure communication. Two main types exist \cite{b1}: local attestation, which authenticates communication between trusted environments on the same processor, and remote attestation, which proves a TEE’s integrity to an external verifiers via signed evidence validated by a trusted authority (e.g., Intel trusted authority).

\subsubsection{Confidential containers }
Kubernetes (k8s) is an open-source container orchestration platform that automates the deployment and management of containerized applications. 
A traditional k8s deployment relies on a Trusted Computing Base (TCB) comprising the host OS and hypervisor, an extensive trust assumption that creates  major security risks \cite{valdez2024crossing}.
\\
The concept of confidential k8s refers to the integration of Kubernetes with confidential computing technologies such as Intel TDX or AMD SEV. Its goal is to run workloads within TEE-enabled systems, specifically Confidential Virtual Machines (CVMs), ensuring that data and code remain protected even from privileged system components. A confidential container platform is designed based on its threat model, adversary capabilities, and the scope of isolation it provides---whether at the container, pod, or cluster level---with the objective of maintaining a minimal Trusted Computing Base.

\section{Methodology and proposed solution}
This section outlines the methodology used to address the identified bottlenecks. We begin by analyzing how the current IT-Wallet design limits self-sovereignty. Next, we describe the main characteristics and architecture of the proposed solution, detailing how participants can securely communicate, present, and validate credentials within an SSI ecosystem.

\subsection{Problems} 
 As described in Section \ref{Introduction}, each participant in the IT-Wallet system is associated with an entity configuration that must be verified. This verification process relies on a PKI-based trust federation, which is  inherently  not compatible with SSI properties. In SSI, decentralization and continuous availability are essential. However, in the IT-Wallet model, the unavailability of even a single trust-mark-status endpoint can prevent verifiers from confirming credential validity.

A valid trust-mark must be issued by an issuer referenced in the trust-mark-issuers claim of the trust anchor or one of its subordinate's entity configurations. Although the trust chain of a credential can provide a certain level of non-repudiation---ensuring traceability even if its provider ceases to exist---the core issue lies in verifying the provider’s status. Specifically, relying on the issuance signature as proof of validity is unsafe, since the status depends on certificate rotation, expiration, and potential key disclosure. Consequently, a stolen private key could still be used to issue fraudulent credentials, meaning that the current model does not guarantee true non-repudiation.

The OpenID federation adopted by the IT-Wallet is built upon the trust assumptions of traditional centralized PKI systems. Most endpoint responses are returned as plain JSON web objects without encryption or digital signatures, which complicates the creation of offline proofs for such interactions. This lack of signed responses is intentional, designed to enhance efficiency and reduce computational overhead by limiting cryptographic operations to cases requiring confidentiality or integrity. According to the OpenID specifications, traditional status web requests to trusted endpoints do not require signing. However, this introduces a verification gap if we want to make these requests attestable: proving that a specific HTTP request and response occurred requires additional mechanisms, such as zero-knowledge proofs of computation or signature, TEEs, or a combination of both.

Additionally, SSI is designed to allow verifiers to validate credentials directly, without relying on third parties or centralized services. In contrast, the IT-Wallet requires interaction with multiple central providers, restricts relying parties to those registered within its system, and limits the use of user credentials to predefined procedures determined by the IT-Wallet providers.

\subsection{Proposed architecture}
We propose a mechanism that leverages IT-Wallet credentials and their existing issuance and verification processes to enable their use within an SSI ecosystem. This approach allows users to retain full control over their credentials, exercising complete autonomy and self-sovereignty, while still being able to prove their credentials validity in accordance with the original IT-Wallet design.

To achieve these objectives, several assumptions are made regarding the IT-Wallet:\\
a) The IT-Wallet provides functionality to export the SD-JWT-VC from its internal storage to an SSI-Wallet. In the current IO app implementation, this feature is neither available nor permitted for security reasons---mainly identity theft and impersonation, which also implies that IT-Wallet credentials are not portable. Although, eIDAS 2.0 emphasizes on interoperability between different Member States solutions, suggesting such feature should be made available in future versions. 
\\b) Exported credentials are assumed not to be time-sensitive to the second---that is, a credential verified as valid at time $t_1$ remains valid for a certain period thereafter, such as days or weeks. This assumption excludes short-lived credentials like the Wallet Unit Attestation (WUA), which are intentionally designed without a revocation mechanism and must instead be verified directly with its provider for each exchange.

The proposed architecture assumes that the user possesses both an IT-Wallet and an SGX TEE-enabled SSI-Wallet capable of generating new TEE-attested credentials. The required interactions are illustrated in Fig. \ref{Creation of TEE-attested credentials} and proceed as follows:
 \begin{figure}[htbp]
\centerline{\includegraphics[width=9.2cm]{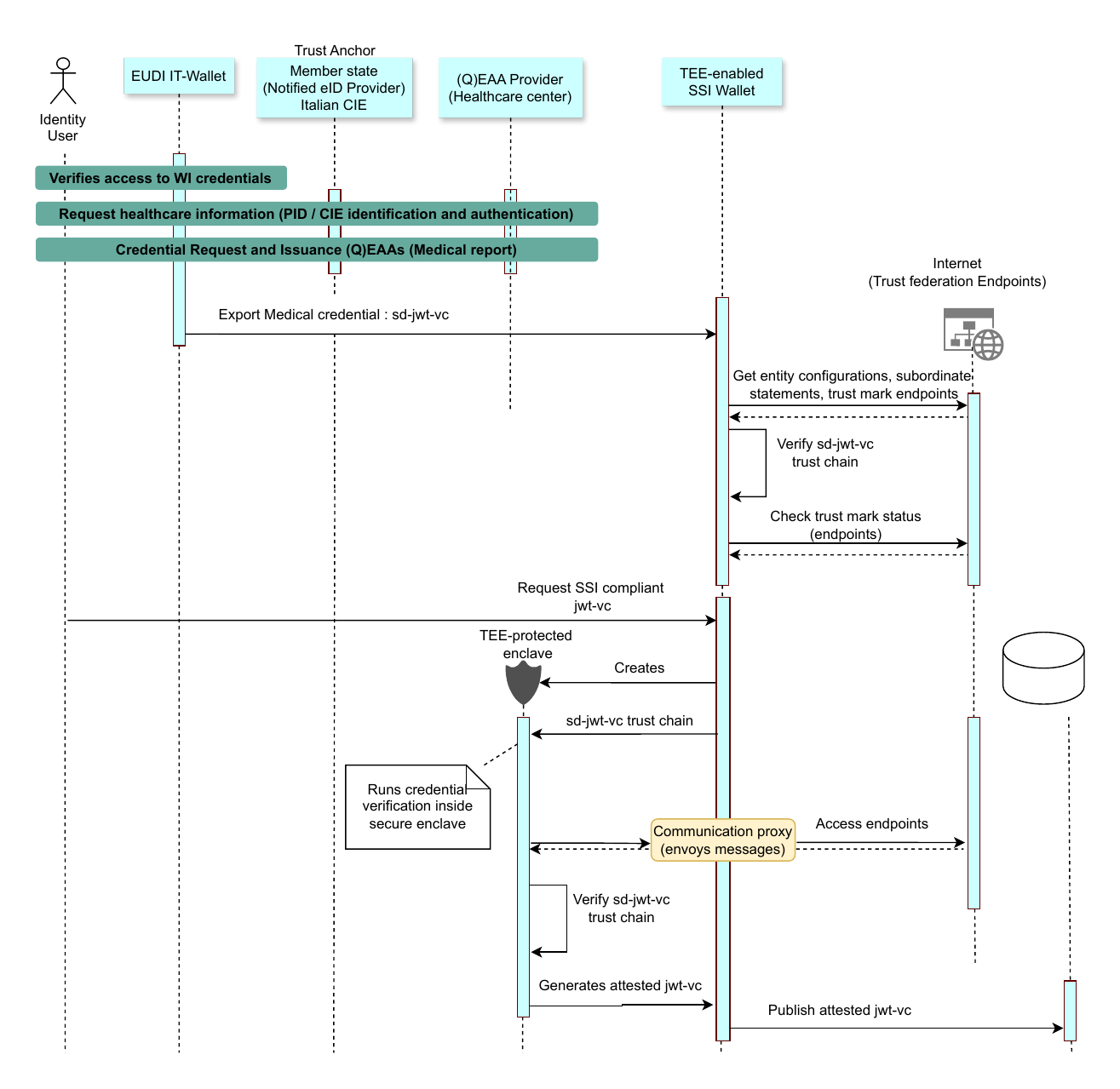}}
\caption{ Creation of TEE-attested credentials }
\label{Creation of TEE-attested credentials}
\end{figure}
\begin{enumerate}[leftmargin=0.6cm]
    \item The user provides his eID credentials to access IT-Wallet instance, then requests credentials from a QEAA provider (e.g. healthcare credential).
    \item The QEAA provider issues a credential in the form of SD-JWT-VC, which is then exported from IT-Wallet to user's TEE-enabled SSI-Wallet. 
    \item The SSI wallet first verifies the credential outside the TEE---saving TEE resources---by retrieving all trust-chain configurations, checking their signatures, and querying each trust-mark-status endpoint to confirm that all qualified providers are active and not revoked.
    \item  The SSI-Wallet initializes a secure enclave that verifies the original SD-JWT-VC along with associated timestamps and predefined policy constraints. Communication between the enclave and external web endpoints is secured using an enclave-enabled TLS library (e.g., mbedTLS-SGX\footnote{{https://github.com/bl4ck5un/mbedtls-SGX}} or OpenSSL-SGX\footnote{{https://github.com/sparkly9399/SGX-OpenSSL}}), where the TLS handshake is initiated within the enclave while the host’s network interface acts solely as a proxy for message transmission.
    
    \item The enclave fetches all required data, validates  trust-marks, verifies credentials, and produces a hardware-attested proof containing endpoint certificates, the original credential, timestamps, and related metadata (using endpoints' certificates as inputs to mitigate DNS-poisoning).
    \item The SSI-Wallet creates a new credential in JWT-VC format, that vouches the original credential (digest) was validated inside an SGX enclave instance. 
    \end{enumerate}
    This credential contains an SGX attestation report (remote attestation), which verifiers can use to verify the enclave identity and the logic that it ran.  
\subsection{Efficient credential verification}
The hardware-attested proof of the new credential must be verifiable in a manner aligned with SSI principles. Given that on-chain verification of such proofs is not efficient, the proof should either be published on-chain in a minimal  form or referenced off-chain ensuring seamless verification between the credential holder and the verifier. Fig. \ref{Succinct publication of the TEE-attested credentials} illustrates how ZK proofs can be integrated to streamline the verification process of the new credential:
\begin{figure}[t]
\centerline{\includegraphics[width=9.5cm]{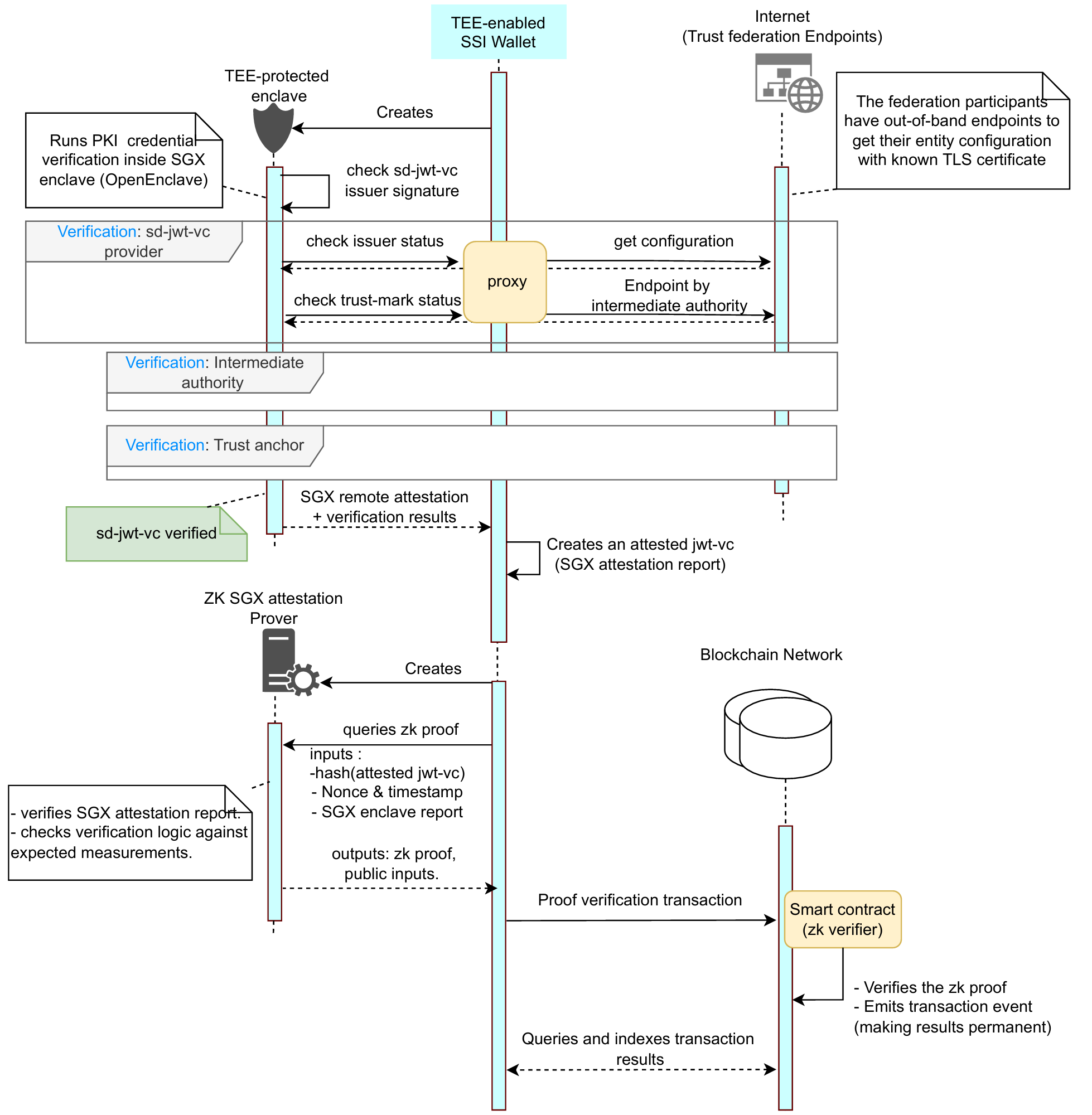}}
\caption{Succinct publication of the TEE-attested credentials }
\label{Succinct publication of the TEE-attested credentials}
\end{figure}
\begin{enumerate}[leftmargin=0.6cm]
    \item  
    During the verification of the original credential within the enclave, the public inputs must include both a hash of the credential and the public certificates of each endpoint. For IT-Wallets, the Trust Anchor’s public key---used to vouch for other endpoints in the federation---must be obtained through out-of-band methods. This approach helps mitigate threats arising from DNS and TLS manipulation.    
    \item Once the enclave verifies all entities in the trust chain and validates the original credential, the SSI-Wallet generates a new attested JWT-VC credential that includes the SGX attestation report and the verification results. This report is signed by the enclave using a provisioned attestation key pair from its SGX platform. Verifiers can validate the SGX quote using a certificate chain rooted in Intel's Provisioning Certification Key (PCK). This process does not compromise execution confidentiality---including from Intel vendor---because the private key associated with the enclave certificate is generated within the enclave and never leaves the hardware-protected CPU space.

    \item 
    The SSI-Wallet employs a ZK SGX attestation prover, which takes as input Intel’s root of trust certificate, the hash of the new credential, and associated metadata. It verifies the SGX attestation evidence while checking that the executed attested logic matches the expected measurements. The prover then produces a ZK proof  and a corresponding ZK verifier smart contract.
    \item The SSI-Wallet then sends a proof verification transaction to the smart contract verifier. This verifier validates the ZK proof and emits a transaction event making these results permanent. 
    The SSI-Wallet submits a proof verification transaction to the smart contract verifier, which validates the ZK proof and emits a transaction event, thereby making the verification results permanent on-chain. Instead of storing the proof and results in the contract’s storage---which incurs a high gas cost---emitting an event makes the same information verifiable at much lower cost.

    \item The SSI-Wallet then either indexes the transaction results---storing the credential hash, transaction hash, and corresponding block for future verification---or subscribes to an indexing service provided by the verification smart contract (most public Remote Procedure Call (RPC) providers acting as indexers listen for events in real time and store them for months or even years). While the latter is simpler, it relies on RPC providers remaining available with full data history and requires trusting them.
  
\end{enumerate}

The new credential can now be directly verified by relaying parties. However, the responsibility of creating the attested credential falls on the identity holder, which is not ideal, since most users typically lack the resources to efficiently run the ZK prover or access TEE-enabled hardware.

 \subsection{Confidential cloud-based proof creation}

To relieve users from the burden of making their credentials available within the SSI ecosystem, a cloud provider can offer this process as a service. However, since the cloud cannot be fully trusted with users’ identification data, we propose a model that leverages confidential Kubernetes or confidential containers to ensure data privacy such as CoCo or edgeless contrast \cite{s1, valdez2024crossing}. Fig. \ref{On-Chain Succinct Proofs of TEE-Attested Credential Validity} illustrates how confidential containers facilitate the verification process of attested credentials:

\begin{figure*}
\centerline{\includegraphics[width=1\textwidth]{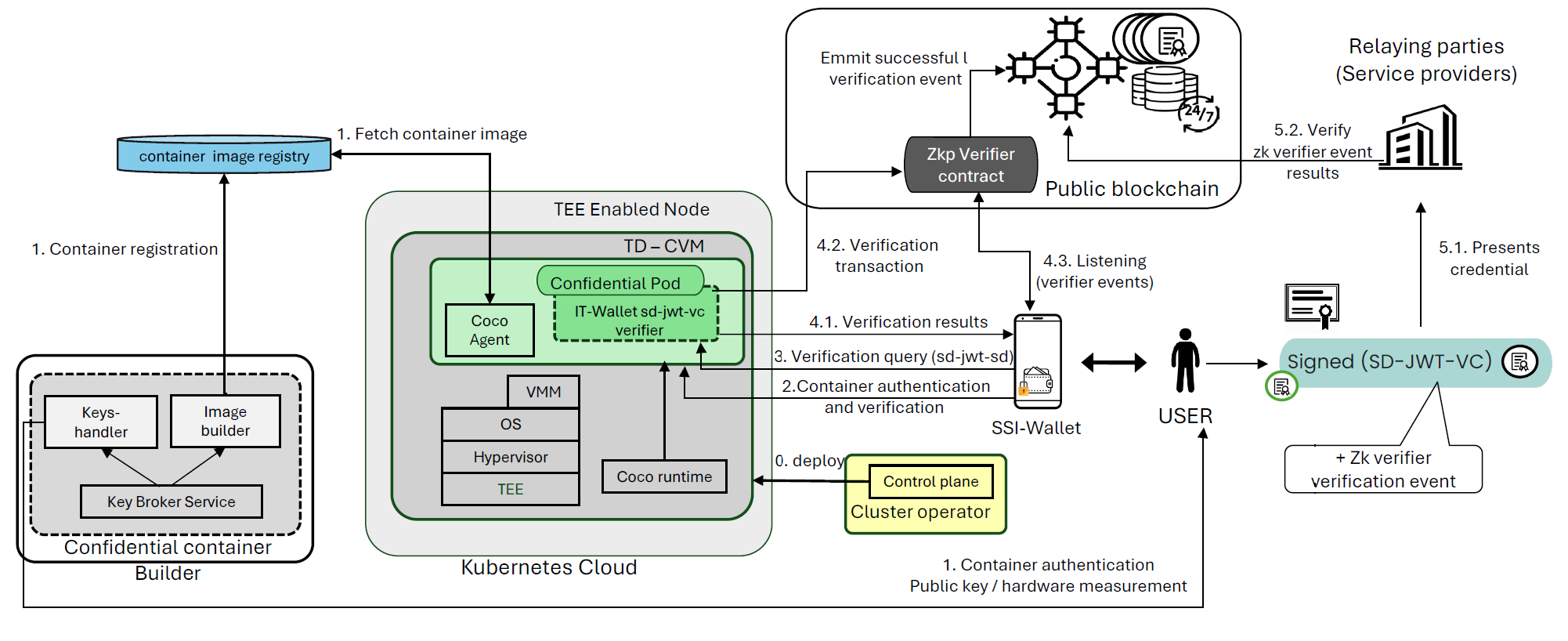}}
\caption{On-Chain Succinct Proofs of TEE-Attested Credential Validity }
\label{On-Chain Succinct Proofs of TEE-Attested Credential Validity}
\end{figure*}

\begin{enumerate}[leftmargin=0.6cm]
    \item The confidential platform is provisioned by a cloud operator running a k8s cluster, which executes the IT-Wallet credential verification workload inside a confidential virtual machine (CVM). This CVM is measured and remotely verifiable, it fetches container images upon initialization. The IT-Wallet credential verification workload must be publicly available to ensure transparency and verifiability.
    \item The SSI-Wallet authenticates the confidential container with the public key of its provider, then verifies that it is running in a TEE-enabled environment with the expected workload image and hardware measurements---hardware reference values must be made available by the cloud provider (they will be attested by TEE attestations). 
    \item Once the confidential IT-Wallet verifier is proven trustworthy and its confidentiality is assured, the SSI-Wallet sends a verification request of the original SD-JWT-VC.
    \item The IT-Wallet credential verifier performs the same verification process described previously---validating the credential and generating a zk proof of the verification results. It then returns the results to the SSI-Wallet and submits a verification transaction to the verifier smart contract.
    \item When a user presents his  SD-JWT-VC to a verifier, it is accompanied with a recorded smart contract event, enabling the verifier to reference and confirm the corresponding successful ZK verification emitted by the smart contract.
\end{enumerate}

The integration of confidential Kubernetes into the verification process offers a practical solution for users seeking to operate their IT-Wallets within an SSI-compliant framework. Because no private identity data are publicly exposed, this approach preserves user privacy and adheres to the data protection principles established by eIDAS 2.0.

\section{Related work}
Most studies focus on combining official eIDs with Web3 technologies---not SSI particularly; \cite{omote2025can} proposes a crypto-asset wallet method that combines eID, one-time wallets and smart contracts by keeping the eID public key secret during transaction verification in public blockchains, preventing the use crypto-assets for laundering while preserving user's privacy.
Authors of \cite{biedermann2024systematisation, baldini2024impact} investigate the incompatibility of OIDC identity protocols under eIDAS 2.0 with web3, deducing that research trends towards developing new trust infrastructures suitable for SSI that further fragment EUDIW and decentralized identity, instead of solutions that combine existing standards for better Web3 interoperability. \cite{biedermann2025aggregating} proposes a novel privacy-preserving authentication protocol making OIDC inherently support Web3 functionalities, a limited unidirectional bridge that limits the root of trust of credentials---this reliance on centralized systems is not SSI-friendly.  
A very relevant solution in \cite{SDIKA-Presentation} provide a cloud-based SSI cloud agent preserving privacy and user control by keeping identity data stored locally. However, performing other identity functions via the cloud provider goes against both cloud and SSI security principles. Others propose alternatives such as relying on ZKP, private standalone wallets, hybrid blockchains (e.g. EBSI) \cite{RamosFernández02092025,fernandez2024evaluation}  to accommodate with eIDAS 2.0. While solutions like PADVA and Janus \cite{szalachowski2019padva, kalka2024comprehensive } rely on TLS notary\footnote{{TLSNotary : {https://tlsnotary.org/}} } protocol to prove data provenance from a TLS session, notarizing publicly available data by executing the TLS communication using privacy-preserving Multi-Party Computation (MPC).

\section{Conclusion and Future Work}
In this work, we have identified several challenges that prevent the EUDIW---and more specifically, the IT-Wallet defined under its directive---from achieving full compliance with SSI principles. To address these limitations, we proposed an architecture that enables IT-Wallet users to leverage their issued credentials within an SSI-compliant environment. Our approach combines TEEs and ZK computations to produce succinct, verifiable proofs without relying on centralized third parties. Furthermore, by incorporating confidential containers, the proposed model enhances usability and accessibility, allowing users of varying technical capabilities to generate proofs securely and efficiently.

A key avenue for future research involves developing prototypes to benchmark the performance of  ZK proofs issuance and assess the costs associated with deploying TEE-based systems. Another important direction is enabling the live verification of short-lived credentials while preserving SSI principles. A promising direction involves the adoption of Merkle tree–based revocation schemes, which could make revocation lists highly available and verifiable at significantly lower computational cost \cite{sitouah2024untraceable, munoz2004certificate}.

\vspace{1cm}
\hspace{-0.4cm}\textbf{Acknowledgment:} N.Sitouah was supported by the Italian Ministry of University and Research (MUR) under NRRP--PNRR codice DOT1316508--and KNOBS srl.


 \bibliographystyle{IEEEtran}
\bibliography{ExportedItems}

\end{document}